# Surface tension of small bubbles and droplets and the cavitation threshold

Mikhail N. Shneider[1,*] and Mikhail Pekker

[1]*Department of Mechanical and Aerospace Engineering, Princeton University, Princeton, NJ USA*

[*]m.n.shneider@gmail.com

## Abstract

In this paper, using a unified approach, estimates are given of the magnitude of the surface tension of water for planar and curved interfaces in the pairwise interaction approximation based on the Lennard-Jones potential. It is shown that the surface tensions of a bubble and droplet have qualitatively different dependences on the curvature of the surface: for the bubble, as the radius of the surface's curvature decreases, the surface tension decreases, whereas it increases on the droplet. The corresponding values of the Tolman corrections are also determined. In addition, it is shown that the dependence of the surface tension on the surface's curvature is important for evaluating the critical negative pressure for the onset of cavitation.

## I. Introduction

The surface tension is an important characteristic of the interface between the contacting media and depends on the intermolecular interaction between them. Attempts to directly calculate the surface tension coefficient have a long history (see, for example, [1, 2]). As was noted in 1949 by Tolman [3], the surface tension coefficient $\sigma$ is not a constant but depends on the curvature of the interface. When the radii of the curvatures are smaller than microns, the surface tension coefficient can differ greatly from the case of a planar interface [4-6].

At present, the Tolman correction to the surface tension coefficient is generally accepted, which is related to the curvature of the surface $R$ [3]:

$$\sigma(R) = \sigma_{\mathrm{T}}(R) = \frac{\sigma_0}{\left(1+\frac{2\delta}{R}\right)} \quad , \tag{1}$$

where $\sigma_0$ is the surface tension of the planar interface and the quantity δ is the Tolman length, which, in general, also depends on the curvature of the interface. Other interpretations of $\sigma(R)$ have also been considered in the literature, for example, the "Lorentz" dependence [5,7,8],

$$\sigma(R) = \sigma_L(R) = \frac{\sigma_0}{\left(1+\left(\frac{\delta}{R}\right)^2\right)}. \tag{2}$$

Many researchers have attempted to directly calculate the Tolman parameter based on molecular dynamics methods [9–13], thermodynamic methods [14–18], and the Gibbs dividing surface method [6, 19]. At the same time, it is generally accepted that, at a given pressure and temperature of the liquid, the Tolman correction is determined only by the curvature of the surface, regardless of the sign of the curvature; that is, it is considered the same for the boundary surface of a droplet and bubble of liquid. However, recent works [6, 17] have found that the dependences of the Tolman parameter on the radius of the curvature are different for the interfaces of a bubble and a droplet, and this difference is not only quantitative but also qualitative. The Tolman correction is very important for the correct evaluation of the critical negative pressure in the liquid for the onset of cavitation [8, 20, 21].

In this paper, a simple analytical method is proposed for calculating the surface tension coefficient on the water-unsaturated vapor interface that takes into account the interaction of water molecules. Also, on the basis of an accepted approach, the surface tension coefficients for bubbles and droplets are calculated depending on the curvatures of their surfaces, and thereby, the corresponding Tolman corrections are determined. Taking into account the data obtained for the surface tension at the interface of the bubble in water, the critical negative pressure at which cavitation begins is calculated. Similar results can be obtained in the framework of the formulated approach for other polar and nonpolar liquids for which the characteristics of the intermolecular interaction potential are known. Also, the results of our simple theory can be used as a test to verify more accurate computer calculations using molecular dynamics methods.

## II. The interaction of molecules in water

Water is a liquid consisting of polar molecules $H_2O$. The interaction of water molecules in various theories and numerical modeling is described by one or more model potentials for intermolecular interaction (contemporary models of intermolecular interactions in water are listed in [22]). A frequently used approximation to describe the interaction of two polar molecules is the Stockmayer potential [23], which depends on the mutual orientation of the constant components of the molecules' dipole moment. In its simplest form, the Stockmayer potential, averaged over all angles of the molecules' mutual orientations, is reduced to the so-called potential of 12-6-3 [24,25]:

$$W = 4W_m \left( \left(\frac{r_0}{r}\right)^{12} - \left(\frac{r_0}{r}\right)^6 - \tilde{\delta}\left(\frac{r_0}{r}\right)^3 \right), \qquad (3)$$

where $W$ is the potential energy and $r$ is the distance between the centers of the two molecules. The third term expresses the contribution of the dipole–dipole interaction, where $\tilde{\delta} = \mu^2/(2W_m r_0^3)$ and μ is the permanent dipole moment of the polar molecule. Generally, $r_0, W_m,$ and $\tilde{\delta}$ are the adjustable parameters, which can be determined from the experiment [24, 25]. When $\tilde{\delta} = 0$, equation (3) is reduced to the classical potential 12-6 of Lenard-Jones [26] for nonpolar molecules:

$$W = 4W_m \left( \left(\frac{r_0}{r}\right)^{12} - \left(\frac{r_0}{r}\right)^6 \right). \qquad (4)$$

However, with an appropriate selection of parameters $r_0$, $W_m$, the Lenard-Jones potential can be used for quantitative estimates of the water characteristics. Various generalized empirical modifications of the Lenard-Jones potential have been proposed (see, for example, recent work [27]), which make it possible to improve the agreement of an experiment with the results of the calculations. In this paper, we confine ourselves to the case of the Lennard-Jones potential (4) and find the corresponding surface tension coefficient. For example, we will use the following set of parameters for the potential (4) $r_0 = 2.65\,A^0$, $W_m = 1.1 \times 10^{-20}$J, obtained from the analysis of the dynamic viscosity of water vapor considered as nonpolar molecules [25].

From the condition that the sum of forces acting on the molecule,

$$F = -\frac{\partial W}{\partial r} = \frac{4W_m}{r_0}\left(12\left(\frac{r_0}{r}\right)^{13} - 6\left(\frac{r_0}{r}\right)^7\right) = 0, \qquad (5)$$

we obtain an estimate of the effective diameter of the molecule:

$$d_m = \alpha r_0 = 2^{1/6} r_0. \qquad (6)$$

The total interaction energy for a test water molecule, summed over the pair of interactions (4) with all molecules, is

$$W = 4W_m \sum_i \left(\left(\frac{r_0}{r_i}\right)^{12} - \left(\frac{r_0}{r_i}\right)^6\right), \qquad (7)$$

where $r_i$ is the distance from the selected test molecule to the i-th one.

Below, we will consider the identical universal algorithm for calculating the surface tension coefficient for planar and curved interfaces, both for positive (droplets) and negative radii of the surface curvatures (micro bubbles).

## III. The surface tension coefficient estimate at a planar interface

Liquid molecules in the bulk and on the surface interact not only with the nearest neighbors but also with markedly distant molecules that are located at distances significantly larger than the size of molecules and intermolecular gaps. Therefore, we will assume that, with respect to each selected molecule, water is a continuous medium with a fixed average density $\rho$. The effective radius of the test molecule's interaction with its nearest neighbors and the density of the molecules are

$$R_m = d_m, \qquad (8)$$

and

$$n_W = \rho/M, \qquad (9)$$

where $d_m$ is determined by formula (6) and $M$ is the mass of the $H_2O$ molecule.

In this case, the total energy of intermolecular pair interactions for a test molecule inside a liquid is

$$W_{in} = 4W_m n_w \int_{d_m = \alpha r_0}^{\infty} 4\pi\left(\left(\frac{r_0}{r}\right)^{12} - \left(\frac{r_0}{r}\right)^6\right) r^2 dr = 4W_m n_w V_m \frac{1}{\alpha^6}\left(\frac{1}{3\alpha^6} - 1\right) = 4W_m n_w V_m I_0, \qquad (10)$$

where $V_m = \frac{4}{3}\pi R_m^3 = \frac{4}{3}\pi d_m^3$ is the effective volume of the interaction of the selected test molecule and $I_0 = \frac{1}{\alpha^6}\left(\frac{1}{3\alpha^6} - 1\right)$.

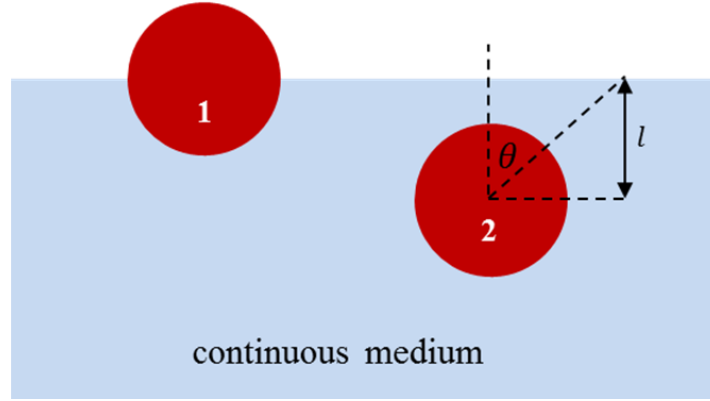

**Fig. 1.** Illustration of the method for calculating the surface tension coefficient of a planar liquid–air interface. 1 - test molecule on the surface of a liquid, $l < d_m$; 2 - test molecule is completely immersed in the liquid, $l \geq d_m$.

We define the coefficient of the surface tension of the liquid $\sigma$ as the difference between the potential energies of the test molecule at the interface and inside the liquid (10), integrated from the interface to infinite liquid depth and divided by the volume $V_m$. In this consideration, the test molecule is represented as a ball with an effective interaction radius $d_m$.

In accordance with the definition of the surface tension coefficient, we have (See Appendix 1)

$$\sigma_0 = 4W_m n_w \left(\int_0^{d_m}(I_{l<d_m} - I_0)dl + \int_{d_m}^{\infty}(I_{l\geq d_m} - I_0)dl\right) = \frac{3}{2\alpha^6}\left(1 - \frac{1}{4\alpha^6}\right)W_m n_w d_m, \quad (11)$$

where, at $y = \cos\theta$, in accordance with Fig. 1,

$$I_{l<d_m} = \frac{2\pi}{V_m}\int_0^{l/d_m} dy \int_{\alpha r_0}^{l/y}\left(\left(\frac{r_0}{r}\right)^{12} - \left(\frac{r_0}{r}\right)^6\right)r^2 dr + \frac{I_0}{2} = \frac{I_0}{2} - \frac{3}{8\alpha^6}\frac{l}{d_m}\left(1 - \frac{2}{5\alpha^6}\right), \quad (12)$$

$$I_{l>d_m} = \frac{2\pi}{V_m}\int_0^1 dy \int_{\alpha r_0}^{l/y}\left(\left(\frac{r_0}{r}\right)^{12} - \left(\frac{r_0}{r}\right)^6\right)r^2 dr + \frac{I_0}{2} = I_0 + \left(\frac{1}{8\alpha^6}\frac{d_m^3}{l^3} - \frac{1}{60\alpha^{12}}\frac{d_m^9}{l^9}\right). \quad (13)$$

Let us find, for example, the value of the surface tension for water at a temperature of 20 °C and atmospheric pressure, $p_0 = 1$ Atm. At these conditions, the density of water is $\rho = 998.23$ kg/m$^3$ and, accordingly, $n_W = 3.34 \cdot 10^{28}$ m$^{-3}$. Substituting in (6) and (11) the values of the accepted parameters of the Lenard-Jones potential from [25], we obtain $\sigma_0 \approx 0.0717$ N/m. The obtained value of $\sigma_0$ differs by no more than 1.5% from the experimentally measured surface tension at the same temperature of 20 °C [28]. Thus, our model, despite its simplicity, gives a good approximation for calculating the coefficient of the surface tension of a liquid.

## VI. Surface tension of an interface with a fixed radius of the curvature

The radius of the interface's curvature can be positive—a droplet of liquid in a gas—or negative— a gas bubble in a liquid, wherein the surface tension of a small droplet or bubble may differ noticeably from the surface tension of a planar interface.

### 1. Surface tension of a bubble interface

Consider, to begin with, the gas–bubble interface in a liquid. Let us determine the surface tension coefficient at the boundary with a gas bubble, similarly to the case of a planar interface. The energy difference between the test molecule near the gas–liquid interface of the bubble and its energy inside the liquid is equal to the interaction energy of the test molecule with the liquid, which would be occupied by the liquid in the bubble (Fig. 2).

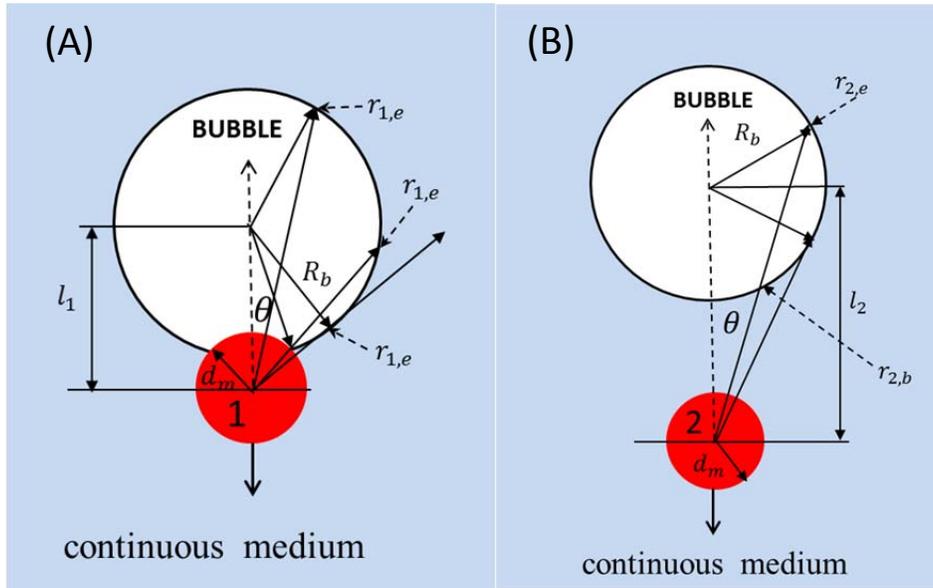

**Fig. 2.** Illustration of the method of calculating the coefficient of the surface tension on the liquid–gas bubble interface. Test molecule 1 is on the boundary of the bubble, and $l_1 < d_m + R_b$ is the distance from the center of the bubble to the center of the molecule. Test molecule 2 is completely immersed in the liquid, and $l_2 > d_m + R_b$ and $r_{1,b}$, $r_{1,e}$ are the beginning and end of the line, respectively, along which integration in (14) for molecule 1 occurs; $r_{2,b}$ and $r_{2,e}$ are the beginning and end of the line, respectively, of integration for molecule 2.

The interaction energy of a probe water molecule $W_b$ located at a distance $l$ from the center of the bubble (if it was occupied by a liquid), in units of $4W_m n_w V_m$, is (see Appendix 2):

$$I_b(l) = \frac{W_b}{4W_m n_w V_m} = \frac{3}{2\alpha^3}\int_{y_0}^{1} dy \int_{r_b\alpha/d_m}^{r_e\alpha/d_m}\left(\frac{1}{x^{10}} - \frac{1}{x^4}\right) dx, \quad y_0 = \frac{\sqrt{l^2 - R_b^2}}{l}. \tag{14}$$

Here $r_e$ and $r_b$ are functions of $l$ and $y = \cos\theta$ (Fig.2). For example, the expressions for $r_e$ and $r_b$:

$$r_b = l\cos\theta - \sqrt{R_b^2 - l^2\sin^2\theta} \qquad (15)$$

$$r_e = l\cos\theta + \sqrt{R_b^2 - l^2\sin^2\theta} \qquad (16)$$

According to the definition of the surface tension coefficient introduced above as the energy difference of the test molecule at the interface and inside the liquid, the surface tension coefficient of the bubble is equal to

$$\sigma_b(R_b) = 4W_m n_w \int_{R_b}^{\infty} I_b(l)\,dl \qquad (17)$$

The dependence of the surface tension coefficient $\sigma_b$ on the bubble radius $R_b$, calculated using formula (17), is shown in Fig. 3 (curve 1). This result is consistent with the assumption of Tolman [3] as well as the results obtained in various approximations by other authors [4-8,14,15,29-31] that the surface tension coefficient of the bubble tends to be zero when the bubble radius decreases. If the bubble radius increases, then $\sigma_b(R_b) \to \sigma_0$ at $R_b \to \infty$, where $\sigma_0$ is the surface tension coefficient for the planar interface. As noted in the introduction, along with the classical interpretation (1) of the surface tension coefficient for small bubbles or drops proposed by Tolman [3], other interpretations of $\sigma(R_b)$ are possible, for example, an approximation (2) (see, for example, [5,7,8]). For comparison, the dependences of $\sigma(R_b)$ corresponding to the approximations (1) and (2) (curves 2 and 3, respectively) also are shown in Fig. 3.

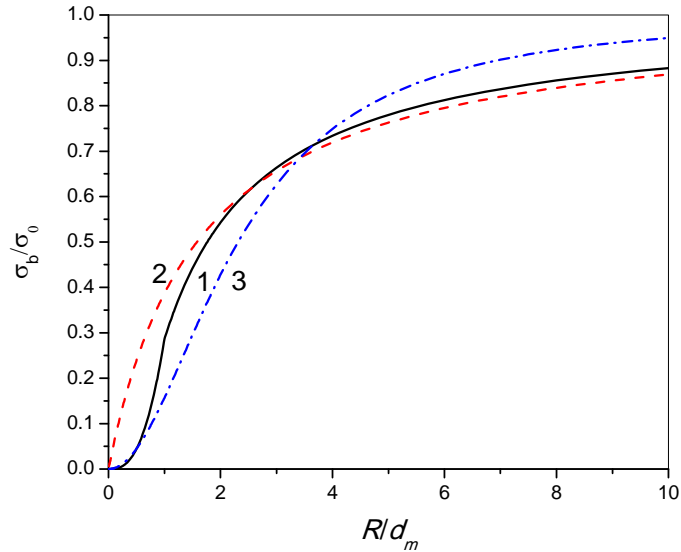

**Fig. 3.** The dependences of the surface tension coefficient $\sigma_b$ on the bubble radius $R_b/d_m$ at water temperature $T = 20\ °C$ and atmospheric pressure. Line 1 corresponds to the results

computed by using formula (11); lines 2 and 3 correspond to the approximation formulas (1) and (2) at assumed values $\delta/d_m = 0.793$ and $\delta/d_m = 2.31$, respectively.

## 2. The surface tension coefficient at the boundary of a droplet in a gas

Another object with a small radius for the surface's curvature is a liquid droplet in a gas. It is natural to expect that the surface tension of the interface of a small droplet also differs from the case of a planar interface. A number of papers stated that, for a droplet, as for a bubble, the Tolman formula (1) is equally valid. Thus, there are calculations confirming this conclusion (see, for example, [12,18]). However, in a number of works, completely opposite results were obtained, showing different qualitative dependences of the surface tension coefficient on the bubble radius and droplet [6,14,15,31]. It was found in these works that, with decreasing droplet size, the surface tension increases rather than decreases, as in the case of a bubble. Fig. 4 shows a diagram of the problem, explaining the integration of a different position of the test molecule in a droplet.

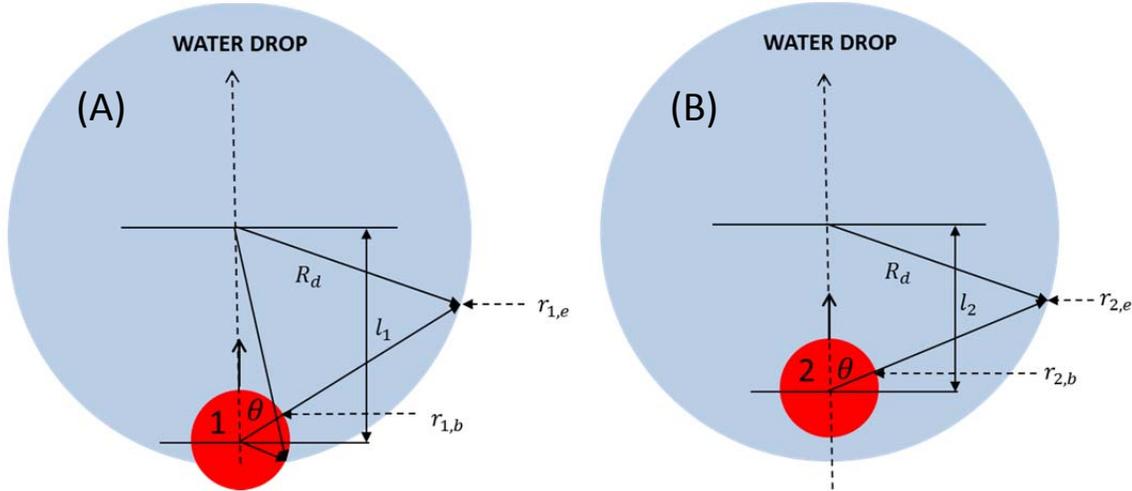

**Fig. 4.** The selected test molecules in a liquid droplet. 1 - a test molecule at the interface, $l_1 > R_d - d_m$; 2 - a test molecule is completely immersed in a droplet, $l_2 \leq R_d - d_m$, $(r_{1,b}, r_{1,e})$ and $(r_{2,b}, r_{2,e})$ are the beginning and end of the line along which integration in (16) takes place for molecules 1 and 2, respectively.

The algorithm for calculating the surface tension coefficient of a droplet with radius $R_d$ is similar to the calculation of the surface tension for a planar interface:

$$\sigma_d(R_d) = 4W_m n_w \left( \int_{R_d-d_m}^{R_d} (I_{l>R_d-d_m} - \tilde{I}_0) dl + \int_0^{R_d-d_m} (I_{l \leq R_d-d_m} - \tilde{I}_0) dl \right), \quad (18)$$

where

$$\tilde{I}_0 = \frac{1}{V_m} \int_{d_m=\alpha r_0}^{R_d} 4\pi \left( \left(\frac{r_0}{r}\right)^{12} - \left(\frac{r_0}{r}\right)^6 \right) r^2 dr = \frac{1}{\alpha^6}\left(\frac{1}{3\alpha^6} - 1\right) - \frac{1}{\alpha^6}\left(\frac{d_m}{R_d}\right)^3 \left(\frac{1}{3\alpha^6}\left(\frac{d_m}{R_d}\right)^6 - 1\right) \quad (19)$$

$$I_{l>R_d-d_m} = \frac{1}{2\alpha^6}\int_1^{y_0}\left(\frac{1}{3\alpha^6}\left(\left(\frac{d_m}{r_b}\right)^9 - \left(\frac{d_m}{r_e}\right)^9\right) - \left(\left(\frac{d_m}{r_b}\right)^3 - \left(\frac{d_m}{r_e}\right)^3\right)\right)dy, \; y_0 = \frac{l^2+d_m^2-R_d^2}{2ld_m} > -1 \quad (20)$$

$$I_{l\le R_d-d_m} = \frac{1}{2\alpha^6}\int_1^{-1}\left(\frac{1}{3\alpha^6}\left(1 - \left(\frac{d_m}{r_e}\right)^9\right) - \left(1 - \left(\frac{d_m}{r_e}\right)^3\right)\right)dy. \quad (21)$$

In formulas (19) - (21) we took into account that $r_b = d_m$, and $r_e = l\cos\theta + \sqrt{R_b^2 - l^2\sin^2\theta}$.

Fig. 5 shows the dependences of the surface tension coefficient on the droplet radius. Since the radius cannot be less than a few characteristic sizes of molecules, we have limited ourselves to the minimum radius of a droplet equal to $R_d = 4d_m$. As the droplet size increases, the calculated surface tension coefficient decreases, thus asymptotically approaching the surface tension coefficient for a planar case. In Fig. 6, it can be seen that the Tolman correction (1) better approximates the surface tension coefficient of a droplet (curve 2) than the "Lorentzian" formula (2) (curve 3).

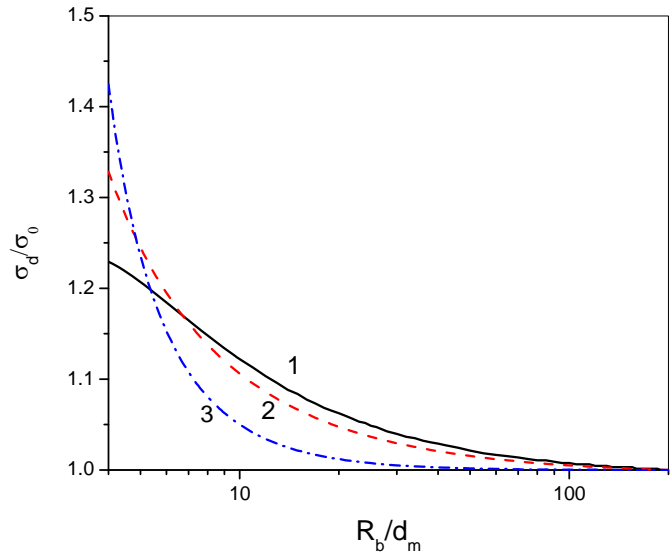

**Fig. 5.** The dependences of the surface tension coefficient $\sigma_d$ on $R_d/d_m$. Line 1 corresponds to our model (16); lines 2 and 3 correspond to the approximation formulas (1) and (2) with $\delta/d_m = -0.5$ and $\delta/d_m = -2.185$, respectively.

## V. Cavitation: critical negative pressure

Surface tension is an important parameter that determines the conditions of cavitation inception. Let us estimate the critical value of negative pressure for the above-found surface tension at the boundary of the bubble $\sigma_b(R_b)$. If the dependence of the surface tension coefficient on the curvature of the surface is not taken into account, it turns out that cavitation in pure water becomes possible only at a negative pressure $|P_-| \ge 200$ MPa. However, experiments show that the critical negative pressure is significantly lower, reaching values $P_{cr} \approx -30$ MPa [32]. This is due to the fact that, for small just emerged bubbles with a radius of several nanometers, the surface tension is substantially less than in the planar case. It was shown in [8,20,21], that the use of Tolman's phenomenological formula (1) or the Lorentzian dependence on the bubble radius (2) for the surface tension coefficient results in theoretical estimates for negative pressure that are much closer to experimental data.

From the approximate linear equation for the state of a compressible fluid [8],

$$p - p_0 = c_s^2 (\rho - \rho_0). \tag{22}$$

it follows, that with reasonable accuracy, the change in the density of molecules at a fixed negative pressure, $P_-$ is

$$n_{W,|P_-|} = n_W \left(1 - \frac{|P_-|}{\rho_0 c_s^2}\right), \tag{23}$$

where $\rho_0$ is the density and $c_s = 1482 \ m/s^2$ is the sound velocity in water at $p = p_0 = 1$ Atm, $T = 20^0$ C, and $|P_-| = 0$. From (23) follows

$$\frac{\sigma_{|P_-|}}{\sigma_0} \approx 1 - \frac{|P_-|}{\rho_0 c_s^2} \quad . \tag{24}$$

The energy of a bubble of radius $R_b$ in a fluid at a constant tensile negative pressure $P_-$ is equal to [32]

$$U_b = -\frac{4\pi}{3} |P_-| R_b^3 + \int_0^{R_b} 8\pi r \sigma_b(r) \, dr \,, \tag{25}$$

where $\sigma_b(r)$ is the surface tension coefficient of a bubble of radius $r$ (Fig. 3), in which the normalization is performed not on $\sigma_0$ but on $\sigma_{|P_-|}$ defined by (24).

The critical radius $R_{cr}$, at which the bubble begins to grow indefinitely, at a given negative pressure is determined from the equality $dU_b/dR_b = 0$. Differentiating (25), we obtain the equation for the critical radius $R_{cr}$:

$$\frac{2\sigma_{|P_-|}(R_{cr})}{|P_-| R_{cr}} = 1 \,. \tag{26}$$

Equation (24) allows us to determine $R_{cr}$ at a given value of $|P_-|$. Then, substituting $R_b = R_{cr}$ in (25), one can find $U_{cr} = U_b(R_{cr})$.

In accordance with [8,20,34,35], taking into account the dependences of the surface tension on negative pressure (24) and the surface curvature radius (Fig. 3), the growth rate per unit volume per unit time for the number of bubbles of a critical radius is

$$\frac{dn_b}{dt} = \Gamma = \frac{3}{4\pi R_{cr}^3} \frac{k_B T}{h} exp\left(-\frac{U_{cr}}{k_B T}\right). \tag{27}$$

The results of the calculations using formula (27) of the dependences of parameter $\Gamma$ on $|P_-|$ are shown in Fig. 6. Line 1 in Fig. 6(A) corresponds to $\sigma_b = \sigma_0$. Lines 1,2, and 3 in Fig. 6(B) show the dependences $\Gamma(|P_-|)$ for the surface tension coefficients corresponding to the lines 1,2, and 3 in Fig. 3.

The criterion for the onset of cavitation is quite subjective. It depends on the choice of the critical growth rate for the number of bubbles in a critical radius per unit time. For example, it can be assumed that cavitation begins to develop intensively when $\Gamma > \Gamma_{cr} \sim 1\ \mu m^{-3} ns^{-3}$, as it may happen in water under the action of electrostriction forces in the vicinity of the pointed electrode as a result of applying a nanosecond high-voltage pulse [8].

Calculations show that, in the case of $\sigma_b = \sigma_0$ (as in the planar case), the absolute value of the critical negative pressure is close to 217 MPa (Fig.6(A), line 1). Yet, even in a planar case, taking into account only negative pressure (24), the absolute value of the critical pressure decreases to 138 MPa ((Fig. 6(A), line 1'). Accounting for the correction (24) and the interface curvature (Fig. 3) noticeably reduces the critical negative pressure to 75–90 MPa (Figure 6 (B), lines 1'–3'), depending on the model chosen for the dependence of the surface tension coefficient on the bubble surface's curvature.

It should be noted that, for conditions when the negative pressure in water is created in other regimes (for example, behind the screw-propeller), the critical negative pressure at which cavitation is initiated may be different. Yet, since $\Gamma(U_{cr}(|P_-|))$ is an exponential function, apparently cavitation begins at a negative pressure that is close enough to the abovementioned values.

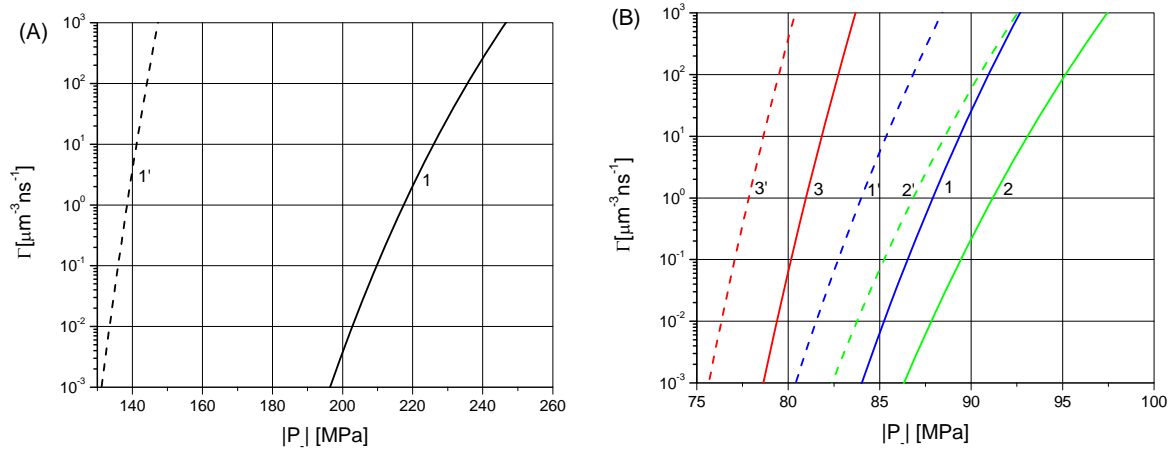

**Fig. 6.** The dependences of $\Gamma$ on the absolute value of the negative pressure for various models of surface tension. (A) Calculation without taking into account the dependence of the surface tension coefficient on the bubble radius. Lines 1 and 1' correspond to $\sigma_b = \sigma_0$ and $\sigma_b = \sigma_0 \left(1 - \frac{|P_-|}{\rho_0 c_s^2}\right)$, respectively. (B) Solid (1,2,3) and dashed (1', 2', 3') lines are calculated with and

without taking into account amendment (22), respectively, and are consistent with the surface tension values 1,2, and 3 shown in Fig. 3.

The obtained dependences of the surface tension coefficient on the bubble radius qualitatively explain the relatively low values of the critical negative pressure at which cavitation occurs. Note that the calculation of Γ from formula (27) does not take into account the effect of saturation for the rate of cavitation bubbles' nucleation considered in [8, 36].

Note that the presence of microscopic defects in water [37] and dissolved gases that are not considered in our analysis can lead to a noticeable decrease in the critical negative pressure. Formally, this can be taken into account by choosing the values of the parameter $\delta$ in the corrections (1) or (2) so that the cavitation starts at the experimental values of negative pressure $P_{cr} \approx -30$ MPa , as was done, for example, in [8,20].

## VI. Conclusions

For the approximate Lenard-Jones potential of intermolecular interaction, a simple formula was obtained for estimating the surface tension of water. The calculated value of the surface tension coefficient coincides with the known experimental data with an accuracy of about 1%.

In the framework of the proposed approach, the dependences of the surface tension on the curvature of the interface surface of small bubbles and liquid droplets were calculated.

It was shown that, for a bubble, with a decrease in the curvature's radius, the surface tension decreases in accordance with the Tolman hypothesis, while for droplets, on the contrary, it increases.

The factors that determine the critical negative pressure for the origin of cavitation were investigated. It was shown that, taking into account the stretching of the fluid as well as the curvature of the surface of the emerging cavitation bubbles significantly reduces the absolute value of the critical negative pressure, getting it closer to the experimental data.

## Literature

1. G. Kirkwood, F.P. Buff, "The Statistical Mechanical Theory of Surface Tension", J. Chem. Phys. **17**, 338 (1949)
2. T. Hill, "Thermodynamics of Small Systems", J. Chem. Phys. **56**, 526 (1952)
3. R.C. Tolman, "The Effect of Droplet Size on Surface Tension", J. Chem. Phys. **17**, 333 (1949)
4. S. L. Bartell, "Tolman's δ, Surface Curvature, Compressibility Effects, and the Free Energy of Drops", J. Phys. Chem. B **105**, 11615 (2001)
5. R. Tsekov, K.W. Stockelhuber, B.V. Toshev, "Disjoining Pressure and Surface Tension of a Small Drop", Langmuir, **16**, 3502 (2000)
6. S. Burian, M. Isaiev, K. Termentzidis, V. Sysoev, L. Bulavin, "Size dependence of the surface tension of a free surface of an isotropic fluid", Phys. Rev. E **95**, 062801 (2017)